\documentclass[a4paper,11pt]{article}
\pdfoutput=1 

\usepackage{jinstpub} 
\usepackage{subfigure}
\usepackage{amssymb}
\usepackage{graphicx}

\title{\boldmath First Results on 3D Pixel Sensors Interconnected to the RD53A Readout Chip
after Irradiation to $1\times$$10^{16}$\,neq cm$^{-2}$ }

\author[a,1]{M. Meschini\note{Corresponding author.}}
\author{on behalf of CMS Tracker,\\}
\author[a]{R. Ceccarelli}
\author[b]{M. Dinardo}
\author[b]{S. Gennai}
\author[b]{L. Moroni}
\author[b]{D. Zuolo}
\author[c]{L. Demaria}
\author[c]{E. Monteil}
\author[d]{L. Gaioni}
\author[e]{A. Messineo}
\author[f]{E. Curr\'as}
\author[f]{J. Duarte}
\author[f]{M. Fern\'andez,}
\author[f]{G. G\'omez}
\author[f]{A. Gar\'ia}
\author[f]{J. Gonz\'alez}
\author[f]{E. Silva}
\author[f]{I. Vila}
\author[g]{G. F. Dalla Betta}    
\author[g]{R. Mendicino}
\author[h]{M. Boscardin}

\affiliation[a]{INFN Firenze, Italy}
\affiliation[b]{INFN and University of Milano Bicocca, Italy}
\affiliation[c]{INFN Torino, Italy}
\affiliation[d]{INFN Pavia and University of Bergamo, Italy}
\affiliation[e]{INFN and University of Pisa, Italy}
\affiliation[f]{IFCA, Spain}
\affiliation[g]{INFN and University of Trento, Italy}
\affiliation[h]{FBK Fondazione Bruno Kessler, Trento, Italy}

\emailAdd{Marco.Meschini@cern.ch}

\abstract{
Results obtained with 3D columnar
pixel sensors bump-bonded to the RD53A prototype 
readout chip are reported. 
The interconnected modules have been tested in a hadron beam 
before and after irradiation 
to a fluence of about $1\times$$10^{16}$\,neq cm$^{-2}$ (1\,MeV equivalent neutrons).
All presented results are part of the CMS R\&D
activities in view of the pixel detector upgrade for the 
High Luminosity phase of the LHC at CERN (HL-LHC). 
A preliminary analysis of the collected data shows hit detection
efficiencies around 97\% measured after proton irradiation.
}


\proceeding{9$^{\text{th}}$ International Workshop on 
Semiconductor Pixel Detectors for Particles and Imaging, 
December 10-14, 2018,  Academia Sinica, Taipei}

\begin{document}
\maketitle
\flushbottom

\section{HL-LHC Requirements and 3D Pixels}
\label{sec:HL-LHC}
Pixel detectors in the innermost layers of the HL-LHC experiments will
have to survive up to a fluence which
can exceed  $2\times$$10^{16}$\,neq cm$^{-2}$, 
while preserving high tracking efficiency \cite{tdr}. 
The total active sensor thickness should 
be about 100 to 150\,$\mu$m in order to keep
both the bias voltage and the power
dissipation after irradiation to a manageable level, while at the same
time allowing for a reasonable amount of collected charge to reach
full hit detection efficiency. 
Moreover, the radiation damage reduces the
effective drift distance of charge carriers because of charge
trapping, so it is not useful, in the case of planar sensors,
to increase the thickness beyond the above limits. 
The 3D pixels, where charge carriers have to travel
distances much lower than the sensor thickness (only 35$\mu$m 
for a $50\times50\,\mu{\rm m}^2$  pixel pitch independently of the
sensor thickness which is the driving parameter for planar pixels), 
are hence very good candidates to satisfy all of the above requirements.

\section{The 3D Columnar Pixel Sensors}
\begin{figure} [htbp]
\centering
\subfigure[$25\times100\,\mu{\rm m}^2$ 2E sensor 
]{
\includegraphics[width=0.404\textwidth,height=5cm,keepaspectratio]{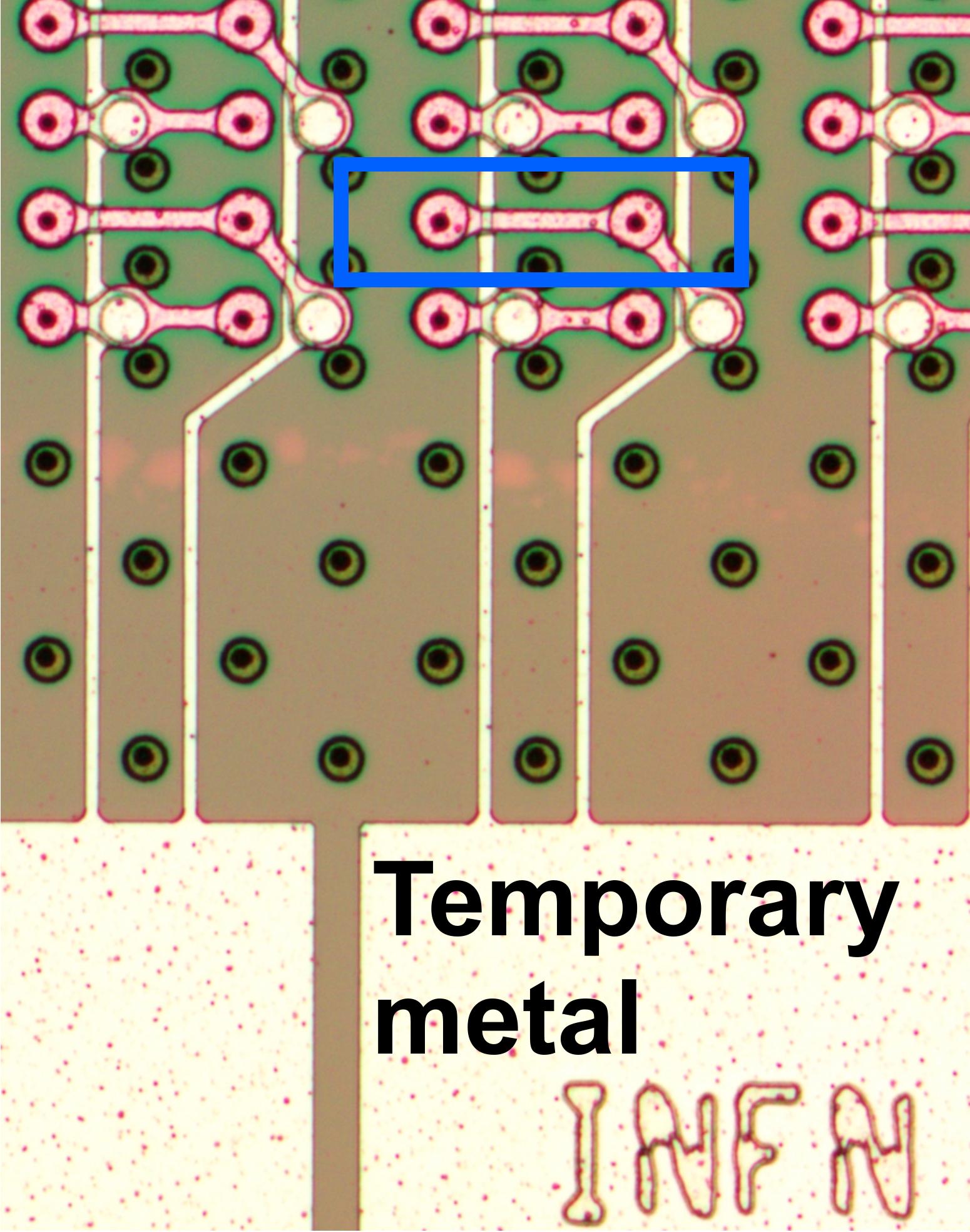}
}
\hspace{1cm}
%
\subfigure[$50\times50\,\mu{\rm m}^2$ sensor %
]{
\includegraphics[width=0.48\textwidth,height=5cm,keepaspectratio]{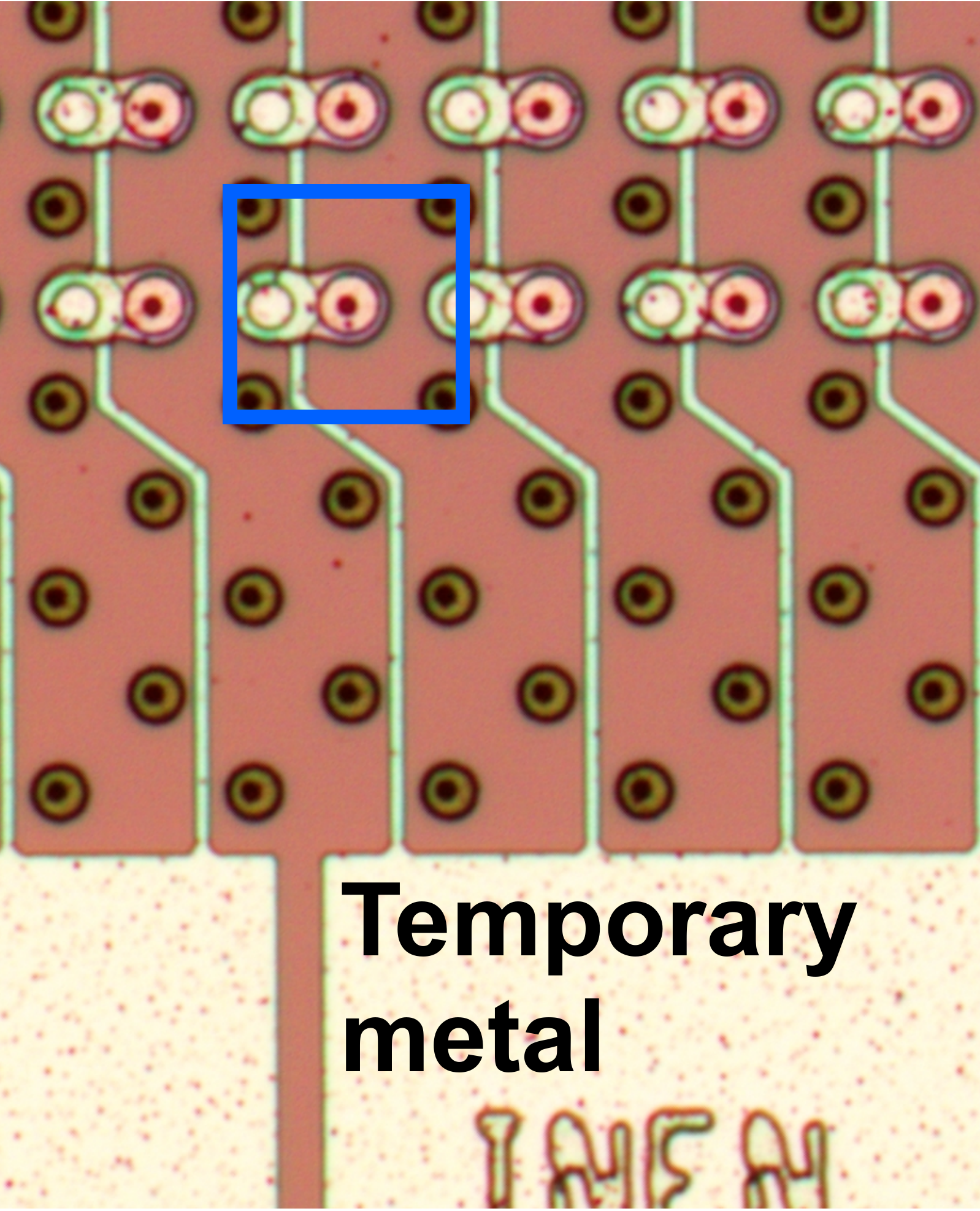}
}
\caption{3D sensor pictures taken with microscope. 
The contact pad for the probe and the vertical metal lines connecting
all pixels are visible in the pictures by the `Temporary metal' label.
The temporary metal layer is used for sensor testing 
at FBK premises and is subsequently removed.  
A 3D single pixel cell on each sensor is highlighted with the blue
frames.
The cell configuration in (a) shows the complexity of the 2E design
and the metal routing to connect readout electrodes to 
the square matrix RD53A bumping pads; the problem is non-existing for
a square pixel cell as in (b).}
\label{pixphoto}
\end{figure}

The 3D sensors \cite{3d} were fabricated at the FBK foundry in Trento; 
they were developed in a collaboration program with INFN
(Istituto Nazionale di Fisica Nucleare, Italy). 
The substrates selected are p-type Si-Si 
Direct Wafer Bond (DWB) or SOI (Silicon On Insulator). 
The handle wafer is 500\,$\mu$m thick low resistivity Czochralski (CZ). 
FBK active devices are implanted on a Float Zone (FZ),
high resistivity (>3000 Ohm cm), 130\,$\mu$m thick wafer. 
Columnar electrodes of both $p^+$ and $n^+$ type are 
etched by Deep Reactive Ion Etching (DRIE) in the
wafer using a top-side only process. 
There are two different pixel cell pitches: $50\times50\,\mu{\rm m}^2$ 
and $25\times100\,\mu{\rm m}^2$, the latter having one or two
collecting electrodes per cell (2E). Two examples of pixel cells are shown 
in figure \ref{pixphoto}. 
%
Both cell sizes are presently under evaluation in CMS 
for the inner layers of the upgrade pixel detectors for HL-LHC. 
After fabrication pixel sensor wafers were processed for 
UBM (Under Bump Metalization), thinned down to $200\,\mu{\rm m}$ total
thickness, diced and bump-bonded to RD53A prototype chips \cite{rd53} 
at IZM (Berlin, Germany). 
 The RD53A chip has 76800 readout channels (400 rows and 192
columns with a bump pad picth of $50\times50\,\mu{\rm m}^2$)
and measures  $20.0\times11.8\,mm^{2}$.
The pixel sensor bonded to the readout chip needs 
eventually to be glued 
and wirebonded onto an adapter card in order to be tested; 
these units will be referred to as modules in the following. 
All results presented here were obtained with the Linear
Front-End in the central zone of RD53A (136 columns wide, from 128 to 263).

\section{Irradiations}
\label{sec:Irr}
\begin{figure} [htbp]
\centering
\subfigure[Modules mounted on the irradiation trays]{
\includegraphics[width=0.404\textwidth]{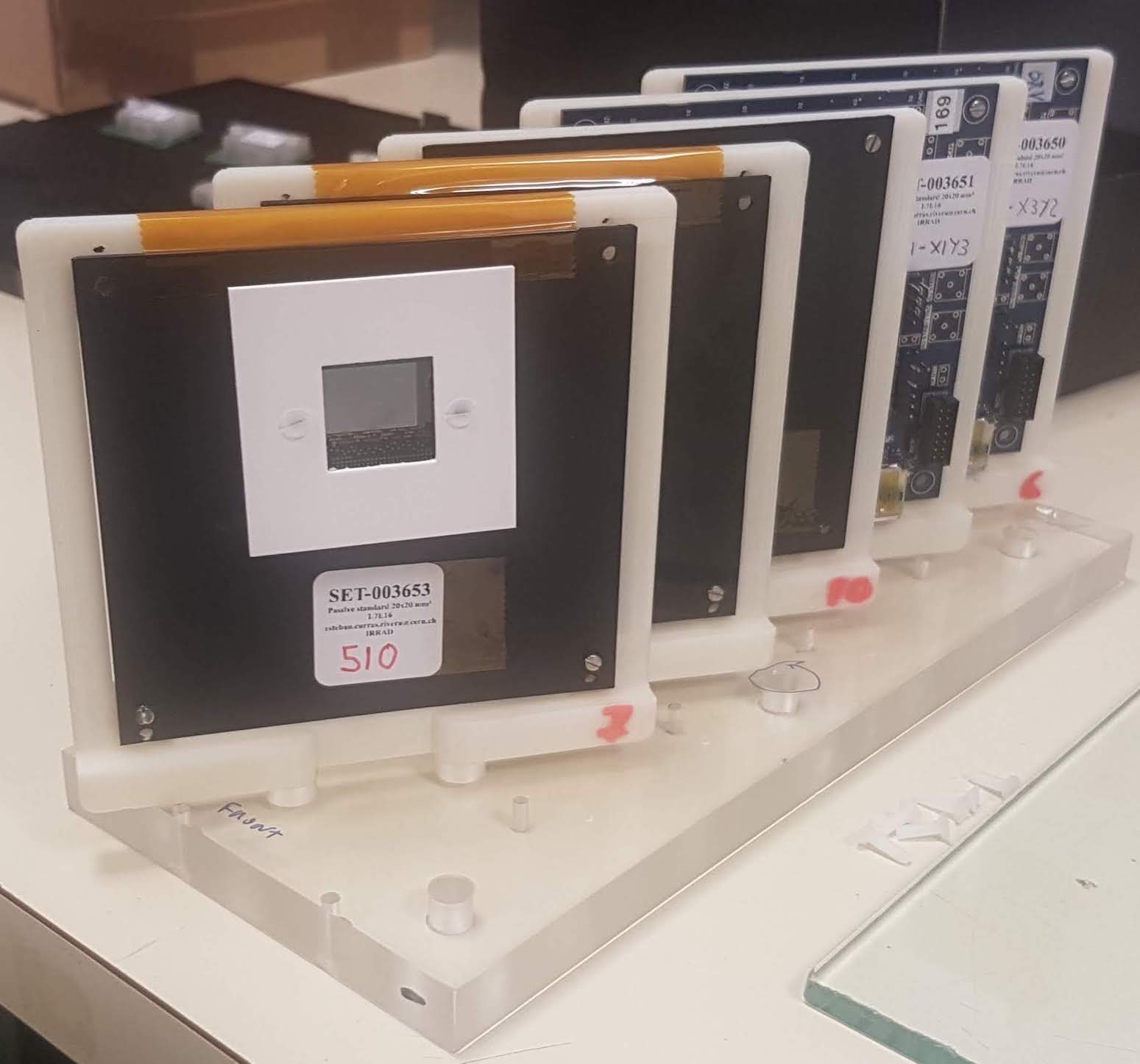}
}
\subfigure[A module after irradiation]{
\includegraphics[width=0.48\textwidth]{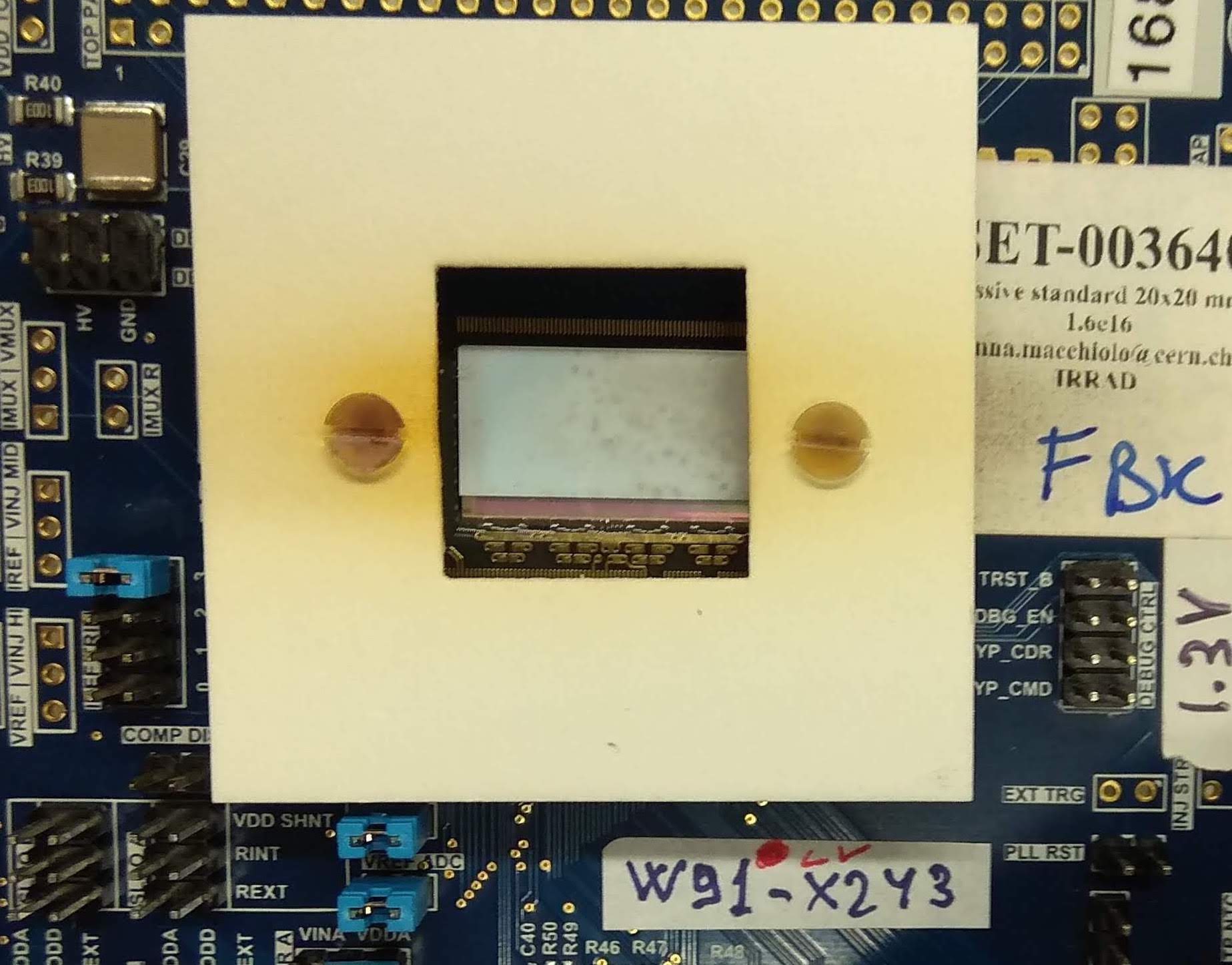}
}
\caption{Tilted modules mounted on the irradiation tray: the two FBK 3D
  modules are the last ones in the stack (a). 
A  module after irradiation (b); the dark brown band on the cardboard frame
  and on the nylon screw heads is 
 due  to the irradiation proton beam passing through the tilted module.}
\label{fig:irr1}
\end{figure}

Irradiations were performed at the CERN IRRAD facility in 2018
in a high intensity 24 GeV/c proton beam, which has a FWHM of 12\,mm in
x and y directions. 
The target fluence was $1\times$$10^{16}$\,neq cm$^{-2}$. 
Modules were tilted on the IRRAD beam at an angle of 55$^\circ$ in order to homogeneously 
irradiate the $20\times12$\,mm$^{2}$ sensor and readout chip area;
the corresponding total ionizing dose was 6\,MGy for 
$1.65\times$$10^{16}$\,protons cm$^{-2}$.
Cross checks were performed in order to establish 
the effective fluence integrated on the modules;
visual inspections after irradiation and data analysis 
are showing that the modules were displaced with respect to 
irradiation beam axis by a few millimeters and that  
the irradiation is less uniform than expected. We estimated that
the nominal fluence was reached in about half of the linear FE.
All results shown here are based on the nominal requested equivalent fluence. 
In figure\,\ref{fig:irr1},  the modules mounted on the adapter cards
are shown. The two CMS modules were labeled w3x3y2 and w91x1y3 to
identify the production wafer and the sensor position on the wafer itself. 
%
%
%
%
\section{Data Taking and Results}
\begin{figure} [htbp]
\centering
\subfigure[Module Threshold]{
\includegraphics[width=0.48\textwidth]{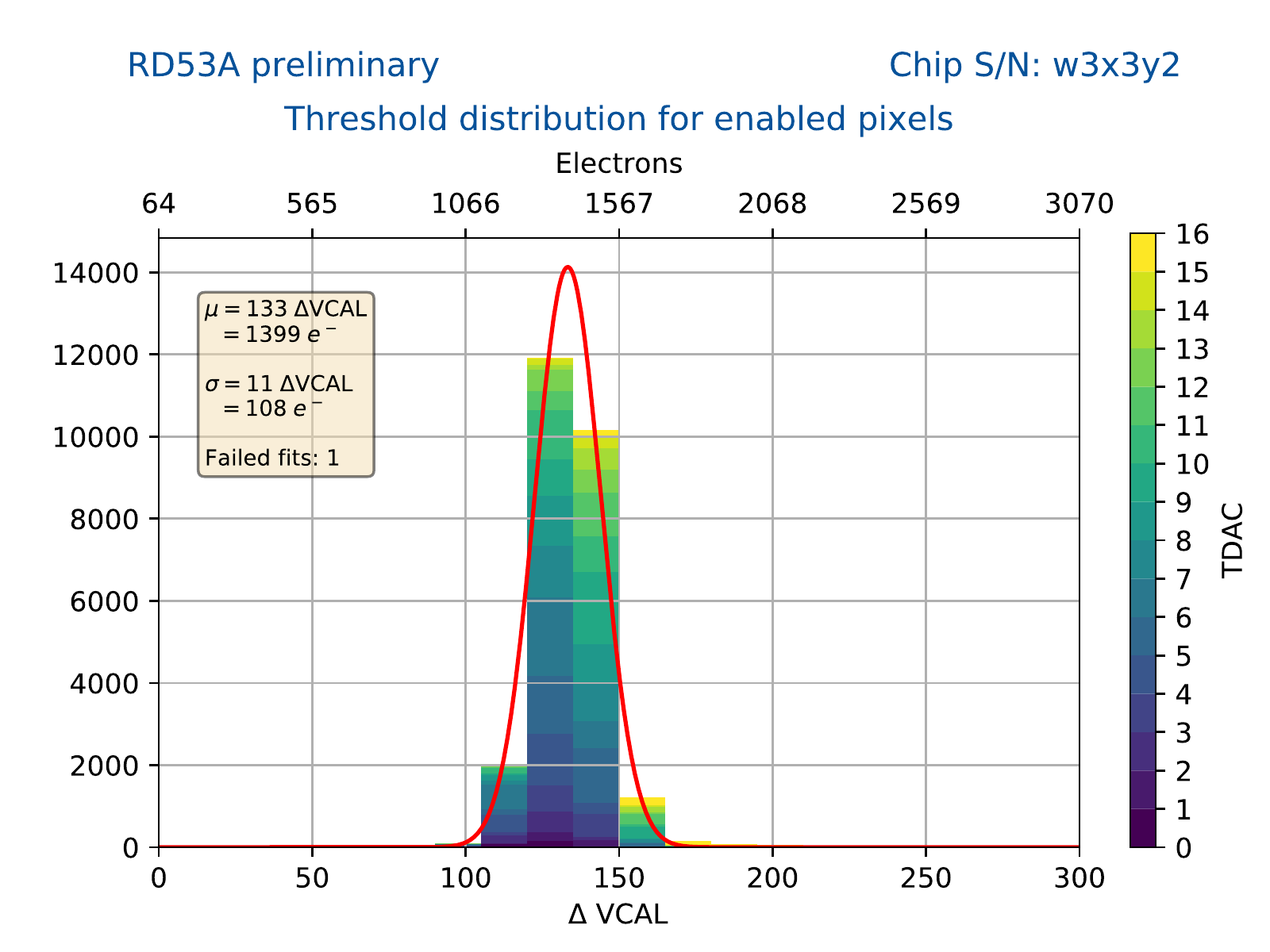}
}
\subfigure[Module noise]{
\includegraphics[width=0.48\textwidth]{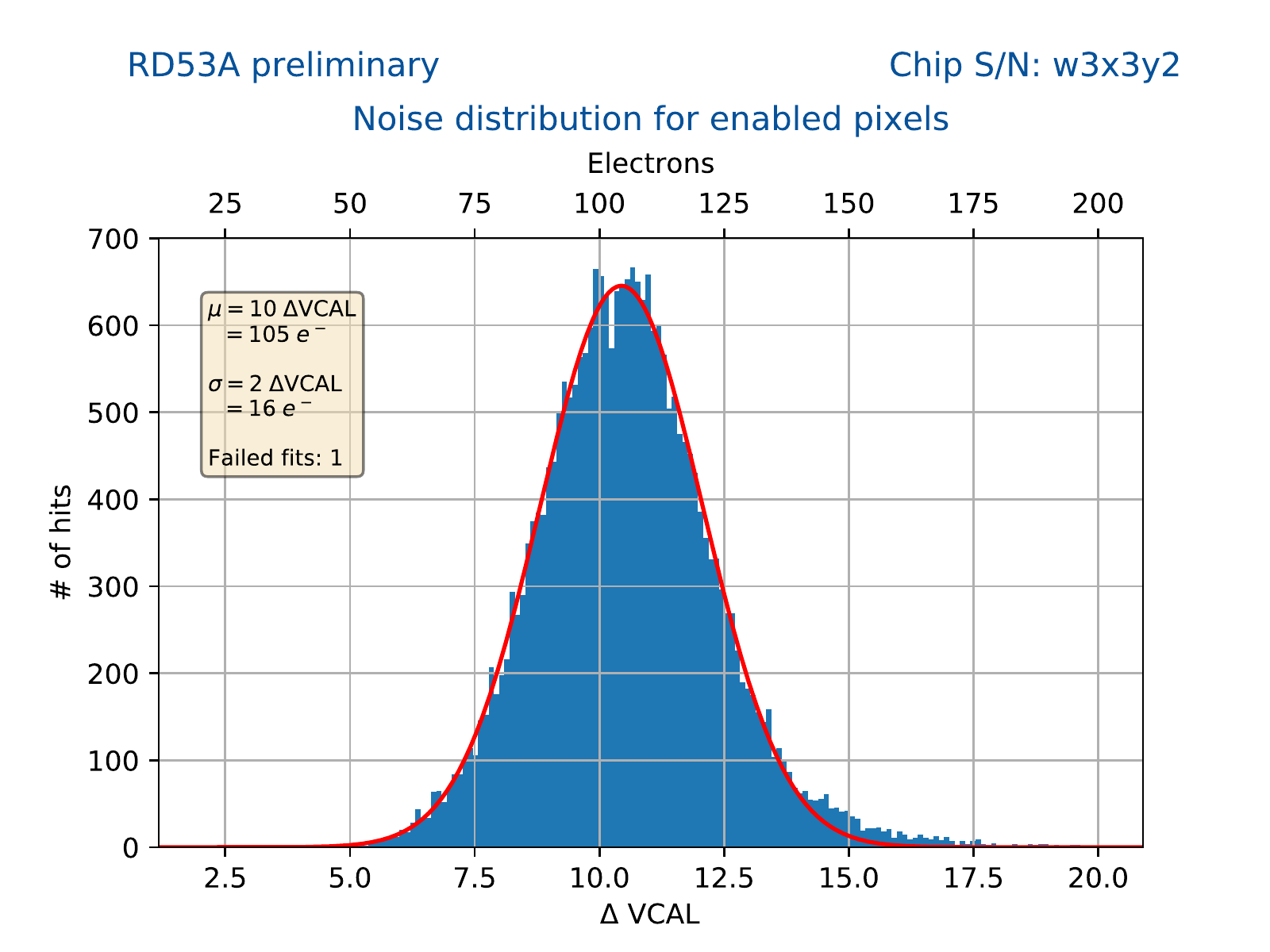}
}
\caption{Signal threshold (a) and noise (b) distributions for 
 module w3x3y2 after irradiation.}
\label{Thr_irr}
\end{figure}

\begin{figure} [htbp]
\centering
\subfigure[Efficiency before irradiation at 3V bias]{
\includegraphics[height=3.5cm, width=0.48\textwidth]{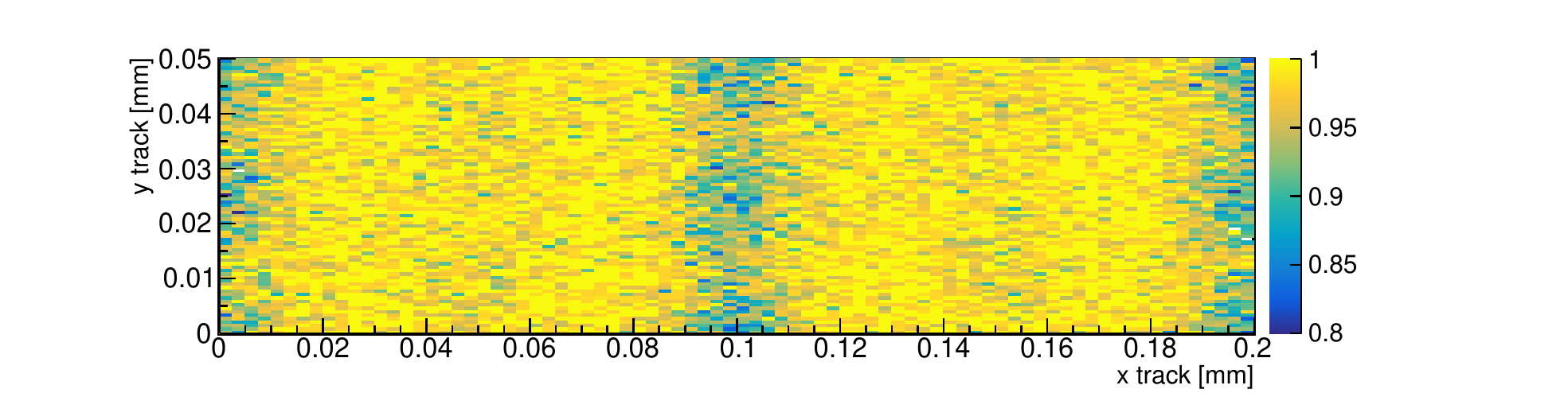}
}
\subfigure[Efficiency after irradiation at 120V bias]{
\includegraphics[height=3.5cm, width=0.48\textwidth]{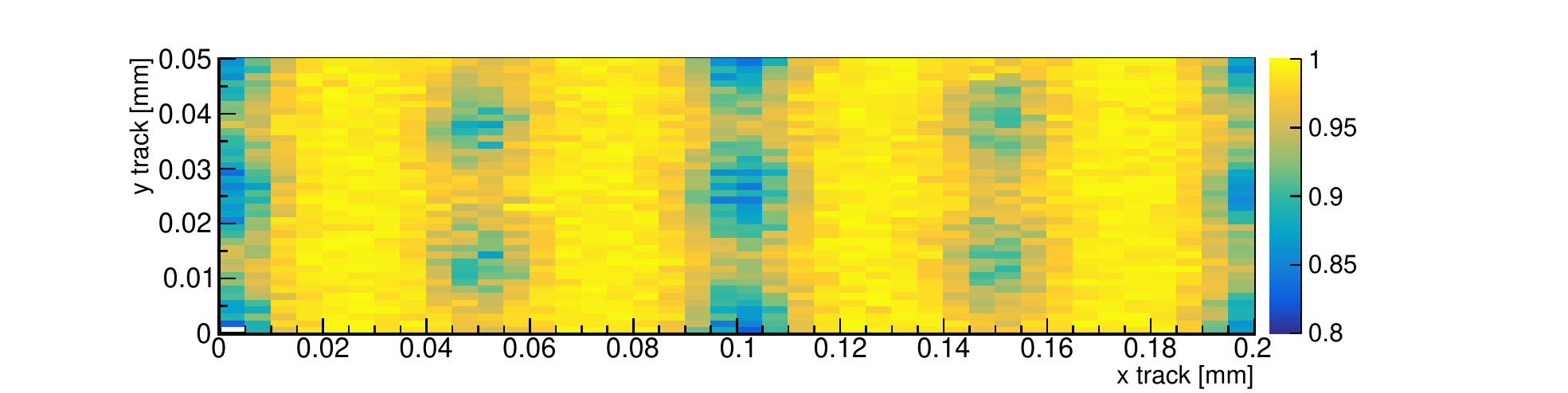}
}
\caption{Hit detection efficiencies  
before  (a) and after irradiation (b)  for the  
$25\times100\,\mu{\rm m}^2$  module w3x3y2.}
\label{25}
\end{figure}

\begin{figure} [htbp]
\centering
\subfigure[Efficiency before irradiation at 15V bias]{
\includegraphics[width=0.48\textwidth]{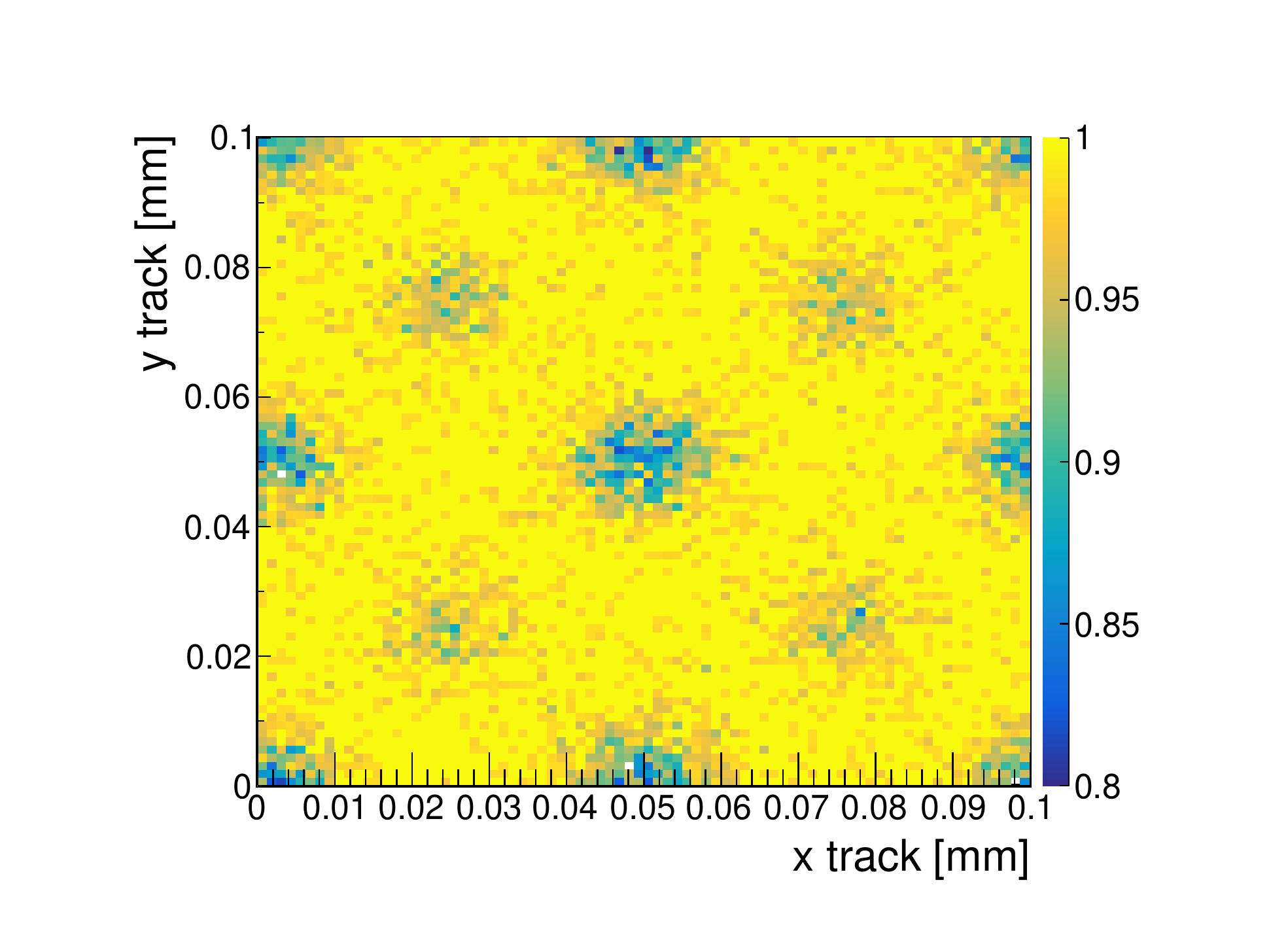}
}
\subfigure[Efficiency after irradiation at 150V bias]{
\includegraphics[width=0.48\textwidth]{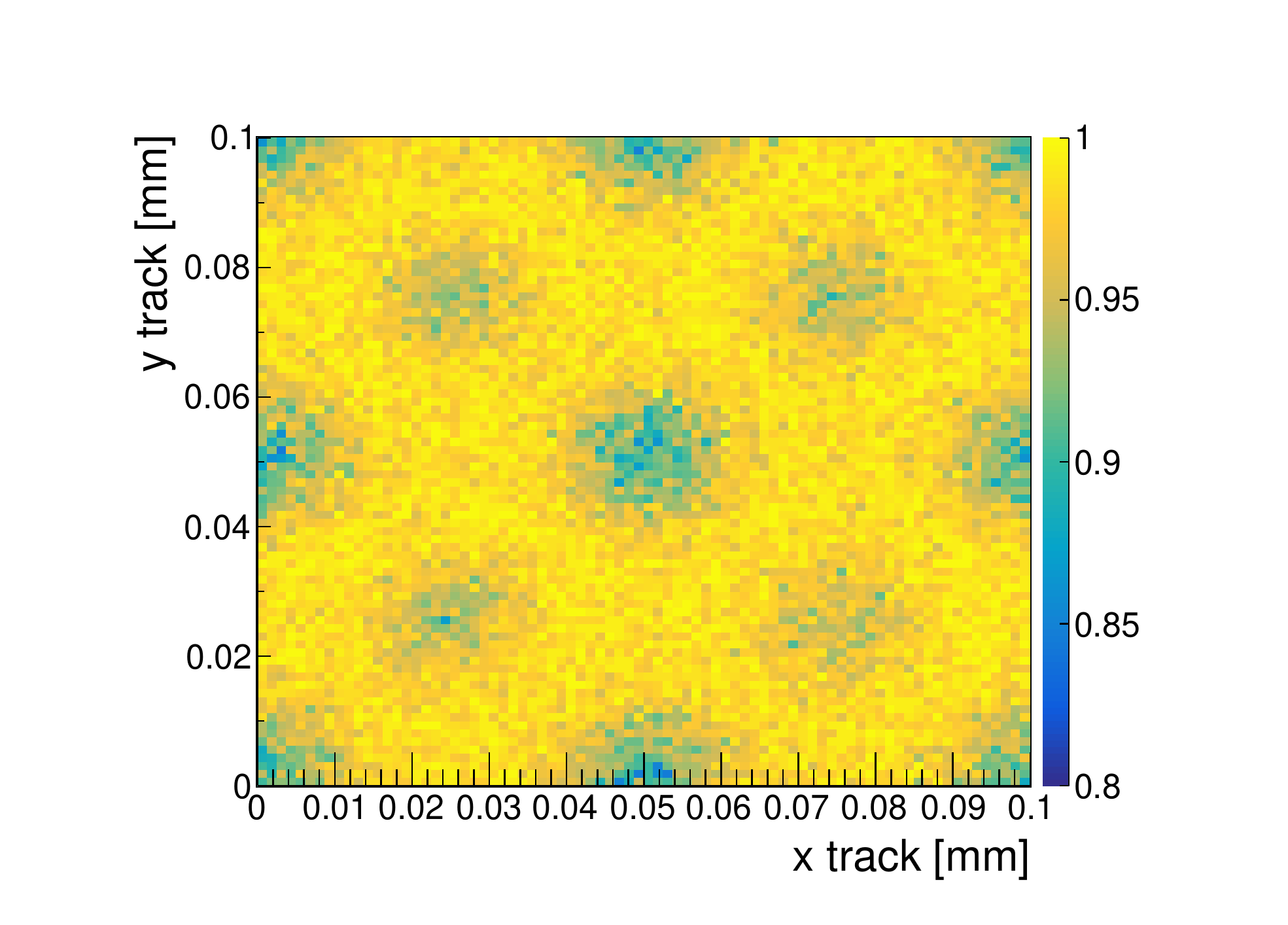}
}
\caption{Hit detection efficiencies 
before  (a) and after irradiation (b) for the
$50\times50\,\mu{\rm m}^2$  module w91x1y3.}
\label{50}
\end{figure}

\begin{figure} [htbp]
\centering
\includegraphics[width=0.65\linewidth]{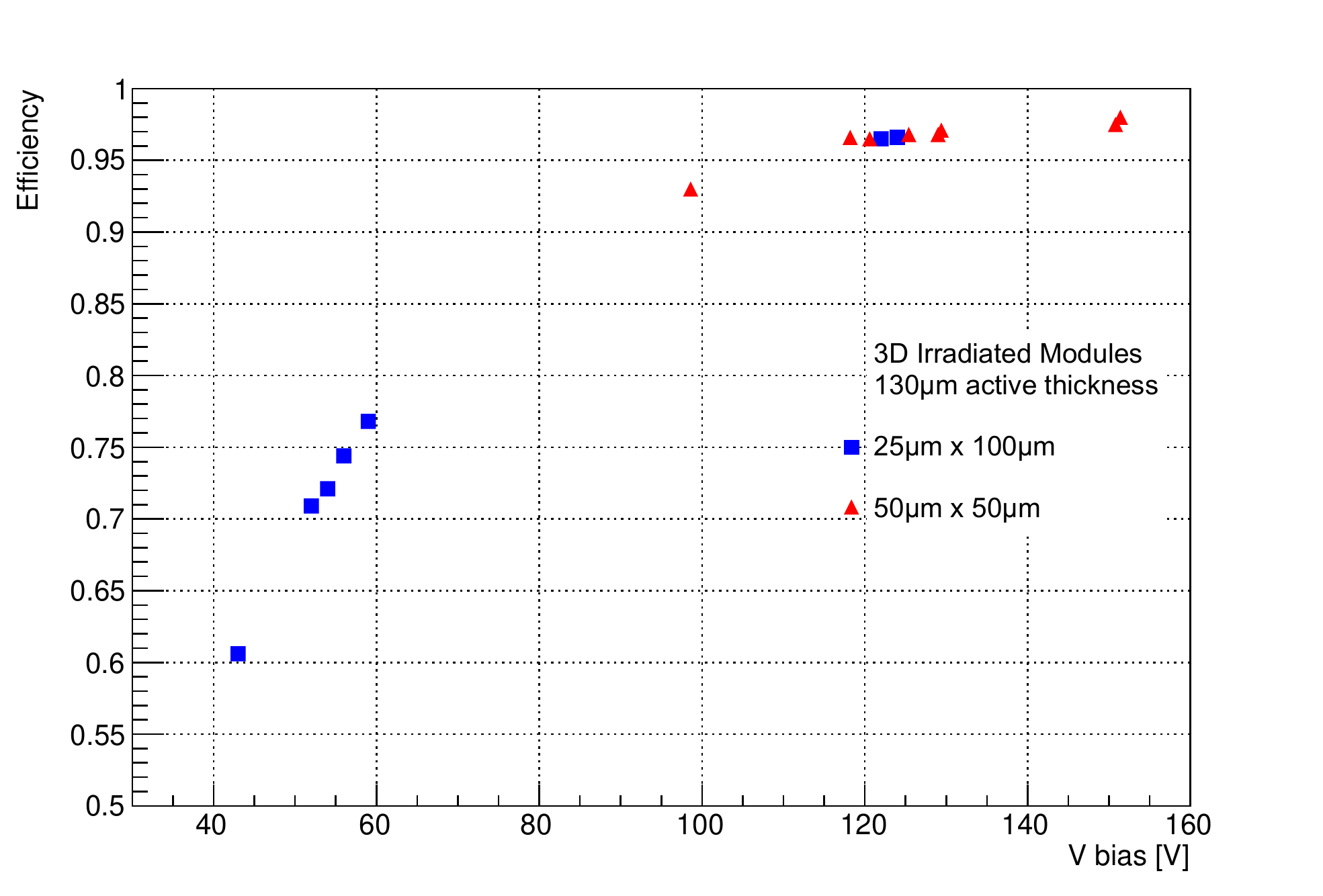}
\caption{Hit detection efficiency vs applied bias voltage}
\label{EffvsHV}
\end{figure}

Hybridized modules were tested in two test beam experiments in
CERN's North Area H6B before and after irradiation in 
July and October 2018. Irradiated modules were kept cold at 
temperatures between  $-20{\rm ^{\circ}C}$ and  $-30{\rm ^{\circ}C}$
using dry ice bricks.  The temperature was monitored via PT1000 sensors 
located close to the backside of the module and via NTC resistors
soldered on the adapter card; these sensors gave consistent measurements. 
The tuning of the readout chip parameters was done on the 
beam-line for all the tested modules, targeting low thresholds and
noise, having at most 1.5\% 
masked pixels because of noisy channels. 
For the irradiated modules the average signal threshold was set 
to about 1400 electrons, with a noise value of 105 electrons  
for non-masked pixels, as shown in figure\,\ref{Thr_irr} for 
a  $25\times100\,\mu{\rm m}^2$ pixel size 
3D module. 
%
The color scale for the threshold distribution represents the 4-bit DAC
value used for the trimming of each individual pixel response.
The 3D modules before irradiation reached hit detection efficiencies above 98.5\%, 
for perpendicular incident tracks, already at moderate HV bias.
In our analysis dead pixels are excluded from the hit efficiency calculation. 
After irradiation, a bias voltage of at least 120V is needed to reach 
high hit efficiency.
A comparison of hit efficiency before and after irradiation is shown
in figure\,\ref{25} for $25\times100\,\mu{\rm m}^2$  
 and figure\,\ref{50} for $50\times50\,\mu{\rm m}^2$ pixel size modules, 
for perpendicular incident tracks. In the efficiency plots the
hits reconstructed over the whole module 
are projected on a 
$2\times2$ pixel cell window to 
put in evidence the sensor geometry and 
the possible effects of the columnar electrodes. 
%
%
%
The geometrical inefficiency due to the columnar electrode
diameter (5\,$\mu{\rm m}$) was estimated to be  around 1.5\%. 
This effect can be greatly reduced by tilting the module on the
beam. 
In non irradiated modules at 34$^\circ$ tilt angle, the hit efficiency
detection reaches 99.3\% for all sensors. 
The hit detection efficiencies as calculated in our data analysis 
for different runs are reported in Table\,\ref{eff}. 
\begin{table}[tbp]
\centering
\caption{\label{eff} Hit detection efficiency summary table.}
\smallskip
\begin{tabular}{|l|c|c|}
\hline
3D Pixel-RD53A Linear FE&$25\times100\,\mu{\rm m}^2$&$50\times50\,\mu{\rm m}^2$\\
\hline
Before irradiation & 97.3\% & 98.6\%\\
After irradiation & 96.6\% & 97.5\%\\
\hline
\end{tabular}
\end{table}
In some selected runs it was possible to take data with
 a different bias voltage in order to verify if the depletion
 region reaches the $p^+$ electrodes delimiting the cell perimeter.
The hit efficiency as a function of the applied bias voltage is
reported in figure\,\ref{EffvsHV} for the available different bias
points.
\section{Conclusions}
The upgrade of the CMS tracker for the HL-LHC will require radiation hard pixel
sensors because of the expected extreme fluences. Thanks to their intrinsic
characteristics, 3D pixels are to be considered as a possible option for
the inner layers of the future trackers. 
Test beam results obtained with FBK 3D pixel sensors bump-bonded to the RD53A
readout chip show high hit efficiencies 
before irradiation already at moderate bias voltages. After irradiation to 
$1\times$$10^{16}$\,neq cm$^{-2}$ the efficiency is 
greater than 96.6\%  at bias voltage of 150V or less. A new batch of 
 3D pixels with stepper technology is in progress at FBK foundry since
December 2018; it will provide very useful data for a deeper study of 
3D sensors and their applications in large quantity. 
More irradiations and beam tests are planned 
to verify at larger scale the performance of these innovative detectors  
for their use in the upgrade trackers. 
\acknowledgments
I wish to warmly thank Mirko Brianzi for his invaluable help in the
construction and wire-bonding of the pixel modules. 
We thank the RD53 Collaboration for the RD53A chip; our results
are not on chip performance but on sensor performance.
This work was supported by the H2020 project AIDA-2020, GA no. 654168.

\end{document}